# The optimum configuration design of a nanostructured thermoelectric device with resonance tunneling

Tong Fu, Jianying Du, Shanhe Su[1], Guozhen Su, Jincan Chen[1]

Department of Physics, Xiamen University, Xiamen 361005, People's Republic of China

**Abstract:** A nanostructured thermoelectric device is designed by connecting a double-barrier resonant tunneling heterostructure to two electron reservoirs. Based on Landauer's equation and Fermi-Dirac statistics, the exact solution of the heat flow is calculated. The maximum power output and efficiency are calculated through the optimizations of several key parameters. The optimum characteristic curve of the performance is obtained. The reasonably working region of the device is determined, the selection criteria of main parameters are provided, and the optimum configuration of the device is drawn. Results show that the heterojunction becomes a perfect energy filter by appropriately regulating the chemical potentials of electron reservoirs and optimally choosing the widths of barrier and quantum well and the nanostructured thermoelectric device with resonance tunneling may obtain simultaneously a large power output and a high efficiency.

[1] Emails: sushanhe@xmu.edu.cn; jcchen@xmu.edu.cn

## 1. Introduction

Nanostructured thermoelectric devices, which offer many advantages for energy conversion with high efficiency, have attracted great interest [1, 2]. There are unexpected harvests of utilizing quantum effects at nanoscales. Paradigmatic examples are quantum dots [3, 4], heterojunctions [5], nanowires [6], and quantum point contacts [7]. By using a quantum dot embedded into a semiconductor nanowire, Josefsson *et al.* demonstrated that thermoelectric conversion efficiency can be close to the thermodynamic limit [8]. A further insight into the operation of the single quantum dot device included the Anderson impurity model and the master equation, which revealed that the efficiency is increased significantly by the cotunneling process [9]. Kuo *et al.* evaluated the properties of the electrical conductance, Seebeck coefficient, and power factor of quantum dot superlattice nanowire arrays by using the tight-binding Hamiltonian combined with the nonequilibrium Green's function method [10]. Nakpathomkun *et al.* compared the performances of three low-dimensional thermoelectric systems, including zero-dimensional quantum dot with a Lorentzian transmission resonance, one-dimensional ballistic conductor, and two-dimensional energy barrier [11].

In recent years, thermoelectric devices with resonant tunneling structures have been extensively studied due to their extraordinary advantages [12-17]. The resonant tunneling effect makes it possible to realize the ballistic transport of electrons [18], and consequently, the performance of a device depends on the energy spectrum of the tunnel. Yamamoto *et al.* studied the Fe/MgO/Fe (001) magnetic tunneling junction by means of the linear-response theory combined with the Landauer-Büttiker approach, where the interface resonant state causes the resonant tunneling and enhances the Seebeck coefficient [19]. A resonant tunneling state also occurs within the forbidden gap of the electron transmittance and creates a giant thermoelectric effect in p and n doped graphene superlattice heterostructures [20]. Castro *et al.* experimentally studied the optical and electronic transport properties of n-type AlSb/GaInAsSb double barrier quantum well resonant tunneling diodes, where a significant resonance current density is observed at room temperature [21].

The energy spectrum of the double barriers-quantum well structure depends on structure parameters [22-25]. One can optimize the performance of thermoelectric devices with double barrier quantum well by controlling the widths of barriers and well. However, by applying the Maxwell-Boltzmann approximation in the directions perpendicular to the direction of the current, most studies assumed that each electron removes an extra average kinetic energy $k_B T$ from a reservoir , where $k_B$ is the Boltzmann constant and $T$ is the temperature of a reservoir [26, 27]. Under certain circumstances, this approximation may underestimate the magnitude of the heat flow. Therefore, for the purpose of revealing the limit of energy conversion, it is necessary to get a stricter analytical solution of the thermodynamic quantities.

In the present study, we will evaluate the performance of the thermoelectric devices with $\mathrm{AL}_l\,\mathrm{Ga}_{1-l}\,\mathrm{As}/\mathrm{GaAs}$ heterojunction by solving the exact solutions of heat fluxes. The rests of this paper are organized as follows: In Sec. 2, the schematic of the thermoelectric device is illustrated. The electronic current density is derived based on Landauer’s formula. Fermi-Dirac (FD) statistics is used to get the exact analytical solutions of the heat fluxes flowing out of reservoirs. In Sec. 3, the transmission probability, net electron current density, power output, and efficiency are evaluated. The optimum selection criteria of main parameters are obtained by maximizing the power output and efficiency. The main conclusions are summarized in Sec. 4.

## 2. Model description

In semiconductor devices, electron transport is usually driven by the temperature and chemical potential differences. In our setup, two electronic reservoirs are connected by a double-barrier resonant tunneling heterostructure, as shown in Fig. 1 (a), which consists of a quantum well of GaAs embedded between two $Al_lGa_{1-l}As$ barriers. The heterostructure allows the electron motion in the $x$ direction completely separated from the $y$ and $z$ directions. Fig. 1(b) indicates the wave vectors of electrons in the momentum space, where $k_x$, $k_y$, and $k_z$ are the wave vectors in the $x$, $y$, and $z$ directions. The electron distribution in a reservoir at temperature $T$ and chemical potential $\mu$ is described by the Fermi-Dirac (FD) distribution function $f\left(E(\vec{k}),\mu,T\right)=\left[1+\exp\left([E(\vec{k})-\mu]/(k_BT)\right)\right]^{-1}$, where $E(\vec{k})=\hbar^2(k_x^2+k_y^2+k_z^2)/2m^*$ expresses the dispersion relation [28] between the wave vector $\vec{k}$ and the kinetic energy of an electron, $\hbar$ is the reduced Planck constant, and $m^*$ is the effective mass of electrons. The hot reservoir is at temperature $T_H$ and chemical potential $\mu_H$, while the cold reservoir is characterized by temperature $T_C$ and chemical potential $\mu_C$.

Only electrons having a kinetic energy $E_x=\hbar^2k_x^{'2}/2m^*$ large enough in the $x$ direction can pass over the barrier, where $k_x^{'}$ is the corresponding wave vector. By assuming that the electron transport process is ballistic, the electronic current density flowing out of a reservoir through the heterostructure is given by the Landauer equation [28, 29]

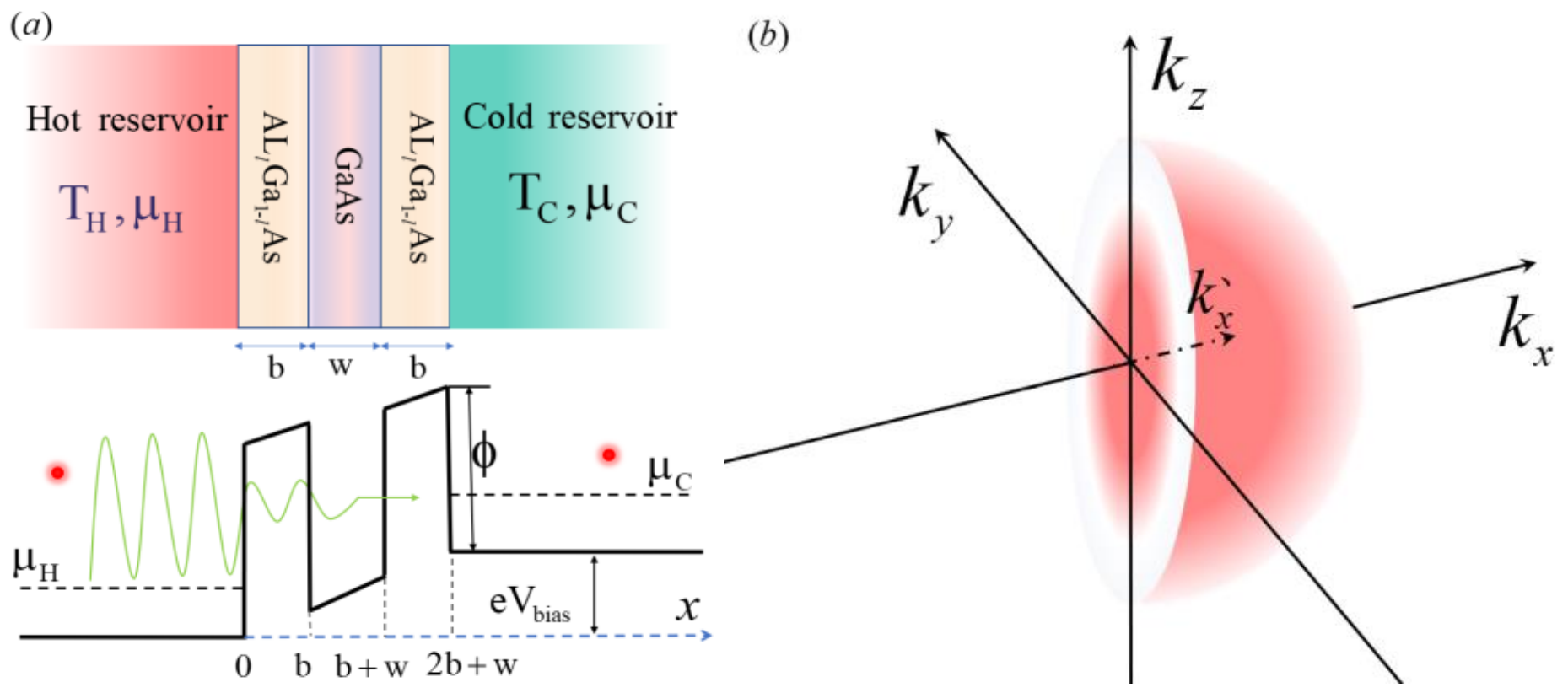

Fig.1. (a) The schematic and band diagrams of the thermoelectric device with two electronic reservoirs connected by a double-barrier resonant tunneling heterostructure. $b$ and $w$ are, respectively, the widths of the layers of $Al_lGa_{1-l}As$ and GaAs. $V_{bias}$ is the bias voltage applied on the heterostructure. The barrier height $\phi$ of $Al_lGa_{1-l}As$ is much higher than the chemical potentials of two reservoirs. (b) The wave vectors for electrons in the momentum space.

$$J = 2e\int_{-\infty}^{\infty}\int_{-\infty}^{\infty}\int_{0}^{\infty} f\left(E\left(\vec{k}\right),\mu,T\right)v\left(k_x\right)\xi(k_x)\frac{dk_x}{2\pi}\frac{dk_y}{2\pi}\frac{dk_z}{2\pi}, \tag{1}$$

where the factor 2 accounts for the degeneracy of electrons and $v\left(k_x\right)=\hbar k_x/m^*$ is the electron velocity in the $x$ direction. $\xi(k_x)$ indicates the transmission probability of an electron tunneling through the heterostructure as a function of $k_x$. It is calculated by the transfer matrix technique [30], which will be summarized in the appendix.

By converting the momentum space into the energy space, Eq. (1) is written as

$$J = \frac{e}{2\pi\hbar}\int_{0}^{\infty} n\left(\mu,T\right)\xi\left(E_x\right)dE_x \tag{2}$$

with

$$n(\mu,T)=\frac{m^{*}k_{B}T}{\pi\hbar^{2}}\log\left[1+\exp\left(-\frac{E_{x}-\mu}{k_{B}T}\right)\right]. \tag{3}$$

The net electronic current density $J_{net}$ is calculated by the difference between the electron current density flowing out of the hot reservoir and that leaving the cold reservoir, i.e.,

$$J_{net}=\frac{e}{2\pi\hbar}\int_{0}^{\infty}\left[n(\mu_{H},T_{H})-n(\mu_{C},T_{C})\right]\xi(E_{x})dE_{x}. \tag{4}$$

According to the first law of thermodynamics, each electron leaving a hot reservoir carries away energy $E-\mu_{H}$ [31, 32], which is the difference between the total energy $E$ of the electron and the chemical potential $\mu_{H}$ of the hot reservoir. An electron from the cold reservoir travelling through the heterojunction will dump the energy that it removes from the cold reservoir plus the work done on it by the chemical potential bias into the hot reservoir, i.e., $E-\mu_{C}-eV$, where $e$ is the elementary charge and $V=(\mu_{C}-\mu_{H})/e$ is the voltage of the device. By combining with Eq. (1), the net heat flux $\dot{Q}_{H}$ flowing out of the hot reservoir is given by [33]

$$\begin{aligned}
\dot{Q}_{H}&=2e\int_{-\infty}^{\infty}\int_{-\infty}^{\infty}\int_{0}^{\infty}(E-\mu_{H})\left[f(E,\mu_{H},T_{H})-f(E,\mu_{C},T_{C})\right]v(k_{x})\xi(k_{x})\frac{dk_{x}}{2\pi}\frac{dk_{y}}{2\pi}\frac{dk_{z}}{2\pi}\\
&=\int_{0}^{\infty}\left\{\frac{m^{*}k_{B}T_{H}}{2\pi^{2}\hbar^{3}}(E_{x}-\mu_{H})\log[1+\exp\left(-\frac{E_{x}-\mu_{H}}{k_{B}T_{H}}\right)]\right.\\
&+\left[\frac{m^{*}(k_{B}T_{H})^{2}}{4\pi^{2}\hbar^{3}}\log[1+\exp(-\frac{E_{x}-\mu_{H}}{k_{B}T_{H}})]^{2}+\frac{m^{*}(k_{B}T_{H})^{2}}{2\pi^{2}\hbar^{3}}\mathrm{Li}_{2}([1+\exp(\frac{E_{x}-\mu_{H}}{k_{B}T_{H}})]^{-1})\right]\\
&-\left\{\frac{m^{*}k_{B}T_{C}}{2\pi^{2}\hbar^{3}}(E_{x}-\mu_{H})\log[1+\exp\left(-\frac{E_{x}-\mu_{C}}{k_{B}T_{C}}\right)]\right.\\
&\left.\left.-\left[\frac{m^{*}(k_{B}T_{C})^{2}}{4\pi^{2}\hbar^{3}}\log[1+\exp(-\frac{E_{x}-\mu_{C}}{k_{B}T_{C}})]^{2}+\frac{m^{*}(k_{B}T_{C})^{2}}{2\pi^{2}\hbar^{3}}\mathrm{Li}_{2}([1+\exp(\frac{E_{x}-\mu_{C}}{k_{B}T_{C}})]^{-1})\right]\right\}\right\}\xi(E_{x})dE_{x}
\end{aligned} \tag{5}$$

with $\mathrm{L}_{n}(t)=\sum_{k=1}^{\infty}\frac{t^{k}}{k^{n}}$ being the polylogarithmic function. The net heat flux $\dot{Q}_{C}$ flowing out of the cold reservoir can be derived by switching the subscripts "$C$" and "$H$". FD statistics is used in Eq. (5) to get the exact analytical solutions of heat fluxes. However, Maxwell-Boltzmann (MB) statistics has been widely adopted in literatures to calculate the energy carried by electrons in the $y$ and $z$ directions [34]. As a result, the energy removed by each electron that leaves a hot reservoir is replaced by $E_x + k_B T - \mu_H$, where $E_x$ is the energy in the $x$ direction and $kT_H$ is the average value of kinetic energy in the $y$ and $z$ directions based on the MB approximation [35]. The electron leaving the cold reservoir may arrive at the hot reservoir and deposits energy $E_x + k_B T - \mu_C - eV$. The net heat flux flowing out of the hot reservoir concerning the relevance of such approximation becomes

$$\dot{Q}_H^{MB} = \frac{1}{2\pi\hbar}\int_0^{\infty}\left[(E_x + k_B T_H - \mu_H)n_H - (E_x + k_B T_C - \mu_H)n_C\right]\xi(E_x)\,dE_x . \quad (6)$$

A similar equation can be derived for the net heat flux $\dot{Q}_C^{MB}$ flowing out of the cold reservoir by using MB statistics. It is worth noting that this approximation may not be accurate under certain circumstances. This problem will be further discussed below. Thus, one has to resort to FD statistics for obtaining the optimal performance of the thermoelectric converter.

By assuming that electrons are transported through an ideal window within a narrow energy range $\delta E$, the transmission probability is given by [36]

$$\xi(E_x) = \begin{cases} 1 & (E_0 - \delta E/2 \le E_x \le E_0 + \delta E/2) \\ 0 & (E_x \le E_0 - \delta E/2; E_x \ge E_0 + \delta E/2) \end{cases}, \quad (7)$$

where $E_0$ is the central energy of the window. Fig. 2 shows the curves of the net heat fluxes out of the hot reservoir varying with $E_0$ based on Eq. (5) (red dashed line) and Eq. (6) (black solid line), which provides a direct comparison between the exact analytical solution and the MB approximation. It is observed from Fig. 2 that when the window is at a range of low energy ( $E_0 \leq 1.87$meV), the heat flows $\dot{Q}_H$ and $\dot{Q}_H^{MB}$ are obviously different. The method based on the MB approximation underestimates the magnitude of the heat flow. The difference between $\dot{Q}_H$ and $\dot{Q}_H^{MB}$ vanishes when $E_0$ is large enough. The reason is that less electrons occupy the high energy levels in the reservoirs. Therefore, both the heat and electron flows approach zero as $E_0$ is very large. Although the MB approximation greatly reduces the complexity of calculation, the result of the heat flow may deviate from the accurate value, especial for the case that $E_0$ is not large enough. In the following discussion, the heat flow calculated by Eq. (5) will be adopted to evaluate the performance of the nanodevice.

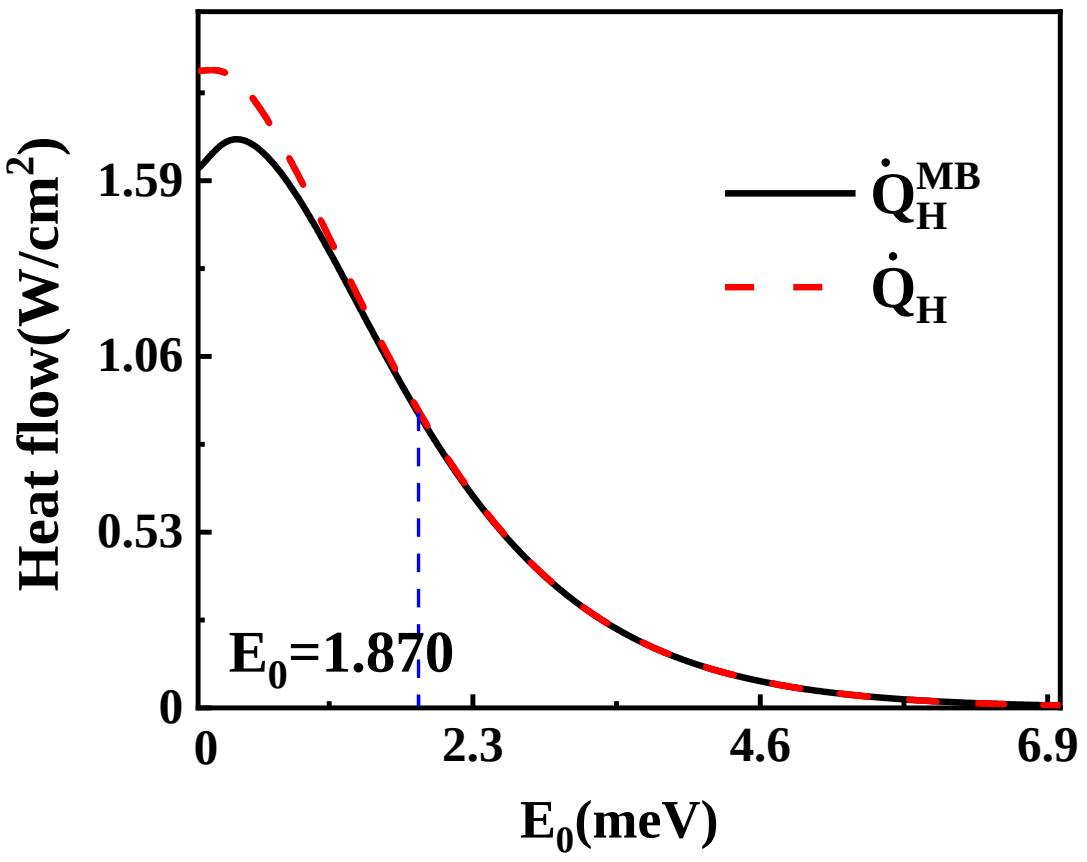

Fig.2. The curves of the net heat fluxes out of the hot reservoir varying with $E_0$ based on the exact solution (red dashed line) and Maxwell-Boltzmann approximation (black solid line ), where $T_H = 10\,\mathrm{K}$, $T_C = 1\mathrm{K}$ , $\mu_H = 0.1\,\mathrm{meV}$, $\mu_C = 200\,\mathrm{meV}$, and $\delta E = 0.0002\mathrm{meV}$. These values are used unless otherwise mentioned specifically in the following discussion. The effective mass of electrons in the hot and cold reservoirs $m^* = 0.1m_e$ with $m_e$ being the free electron mass.

## 3. Results and discussion

For the heterostructure shown in Fig. 1(a), the transfer matrix technique (Appendix A) is used to calculate the transmission probability $\xi(E_x)$. Fig. 3(a) gives the curves of the transmission probability $\xi(E_x)$ as a function of $E_x$ at $V_{bias} = 0V$ for different widths of barrier and well. It is clearly shown that the half peak width of the first resonance peak depends on the width $b$ of the barriers. When $w = 3.5$ nm and $b$ is changed from 3.5 nm (black dash-dotted line) to 4.0 nm (green dashed line), the first resonance peak becomes narrower as its half peak width decreases. On the other hand, the resonant energy $E_{res}$ corresponding to the maximum transmission probability $\xi_{max}$ of the first resonance peak is mainly determined by the width of the well $w$. When $b = 4.0$ nm and $w$ decreases from 3.5 nm (green dashed line) to 3.0 nm (red solid line), the first resonance peak moves to the right with higher energy level. Note that the half peak width of the first resonance peak increases slightly with the decrease of $w$ as well, as indicated by the inserted figure in Fig.3. (a).

In order to reveal the influences of the bias voltage, the maximum transmission probability $\xi_{max}$ of the first resonance peak and its corresponding resonant energy $E_{res}$ varying with $V_{bias}$ are presented in Fig.3. (b), where $b = 4.0$ nm and $w = 3.5$ nm. It is shown that $\xi_{max}$ monotonically decreases with the increase of $V_{bias}$ (black solid line), while $E_{res}$ is a monotonically increasing function of $V_{bias}$.

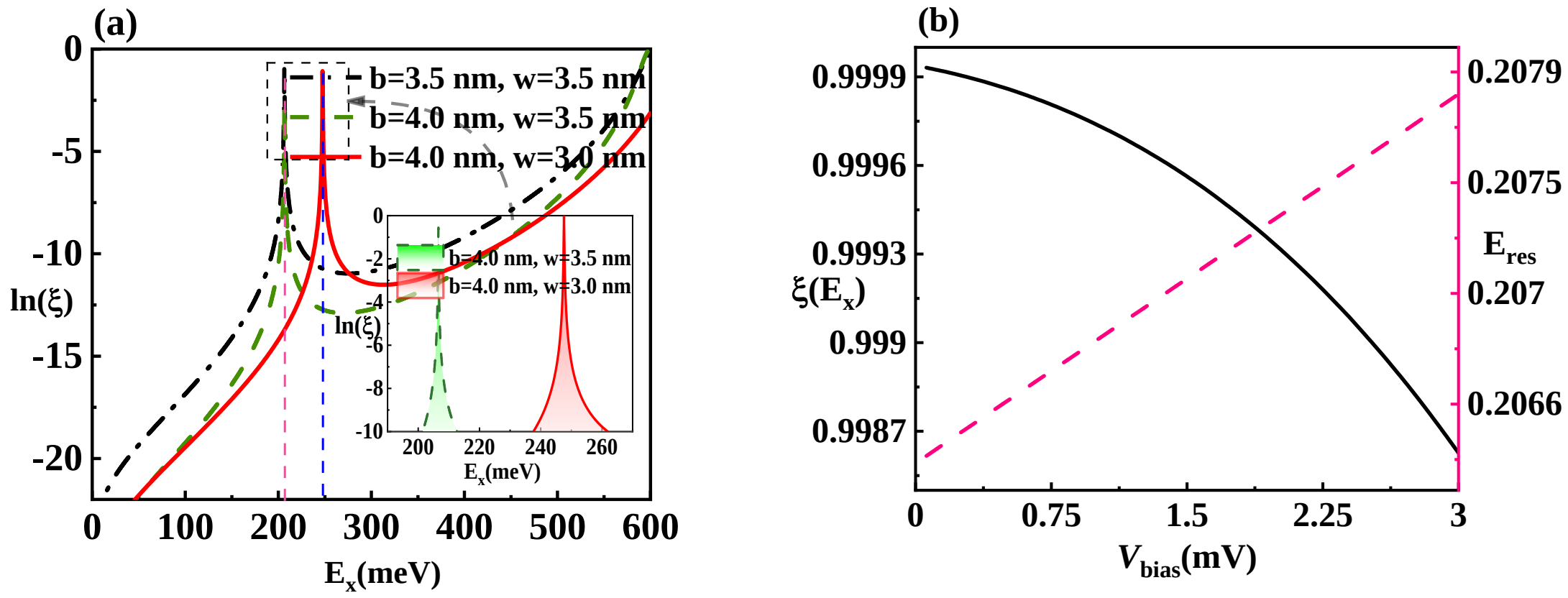

Fig.3. (a) The curves of the transmission probability $\xi(E_x)$ varying with $E_x$ at $\mathrm{V_{bias}} = 0\,\mathrm{mV}$ for different widths of barrier and well as labelled. The inserted figure shows the curves of $\xi(E_x)$ varying with $E_x$ for $b = 4.0\,\mathrm{nm}$ and $w = 3.0\,\mathrm{nm}$ and $3.5\,\mathrm{nm}$, where the range of $E_x$ is from $E_x = 180\,\mathrm{meV}$ to $E_x = 280\,\mathrm{meV}$. For the heterostructure, we choose $\phi$=600 meV and $l = 0.33$. (b) The curves of the maximum transmission probability $\xi_{\max}$ of the first resonance peak and its corresponding resonant energy $E_{res}$ varying with $\mathrm{V_{bias}}$ for $b = 4.0\,\mathrm{nm}$ and $w = 3.5\,\mathrm{nm}$. The left vertical axis shows the value for $\xi_{\max}$, while the corresponding scales of $E_{res}$ is on the right vertical axis.

In the operating regime of the thermoelectric device, the thermodynamic affinity due to the temperature difference of reservoirs drives the electronic flow against the bias voltage $\mathrm{V_{bias}}$. Fig.4 shows the electronic current $J_{net}$ as a function of $\mathrm{V_{bias}}$, where the chemical potential of the cold reservoir $\mu_C = 200\,\mathrm{meV}$. The electronic flow is mainly determined by the first resonance peak, because the second resonance peak appears at the energy level much larger than $E_{res}$. Note that $E_{res}$ increases with $\mathrm{V_{bias}}$, less electrons exists in higher energy levels, leading to the reduction of $J_{net}$. $J_{net}$ at $b = w = 3.5\,\mathrm{nm}$ (black dash-dotted line) is larger than $J_{net}$ at $b = 4.0\,\text{nm}$ and $w = 3.5\text{nm}$ (green solid line). This phenomenon can be explained by two aspects. The half peak width of the first resonance of $\xi(E_x)$ at $b = w = 3.5$ nm is larger than that of $\xi(E_x)$ at $b = 4.0\,\text{nm}$ and $w = 3.5$ nm [Fig. 4(a)]. On the other hand, the first resonance peak moves to higher energy levels as $b$ changes from $b = 3.5$ nm to $b = 4.0\,\text{nm}$, as indicated by the inserted figure in Fig. 4. In the case of $b = 4.0\,\text{nm}$ and $w = 3.0$ nm, $J_{net}$ (red solid line) is quite small compared to the other two cases in Fig. 4, because the transmission probability $\xi(E_x)$ shifts to energy levels with fewer electron occupation numbers. In general, both the shape and the energy range of the first resonance peak influence the electronic flow.

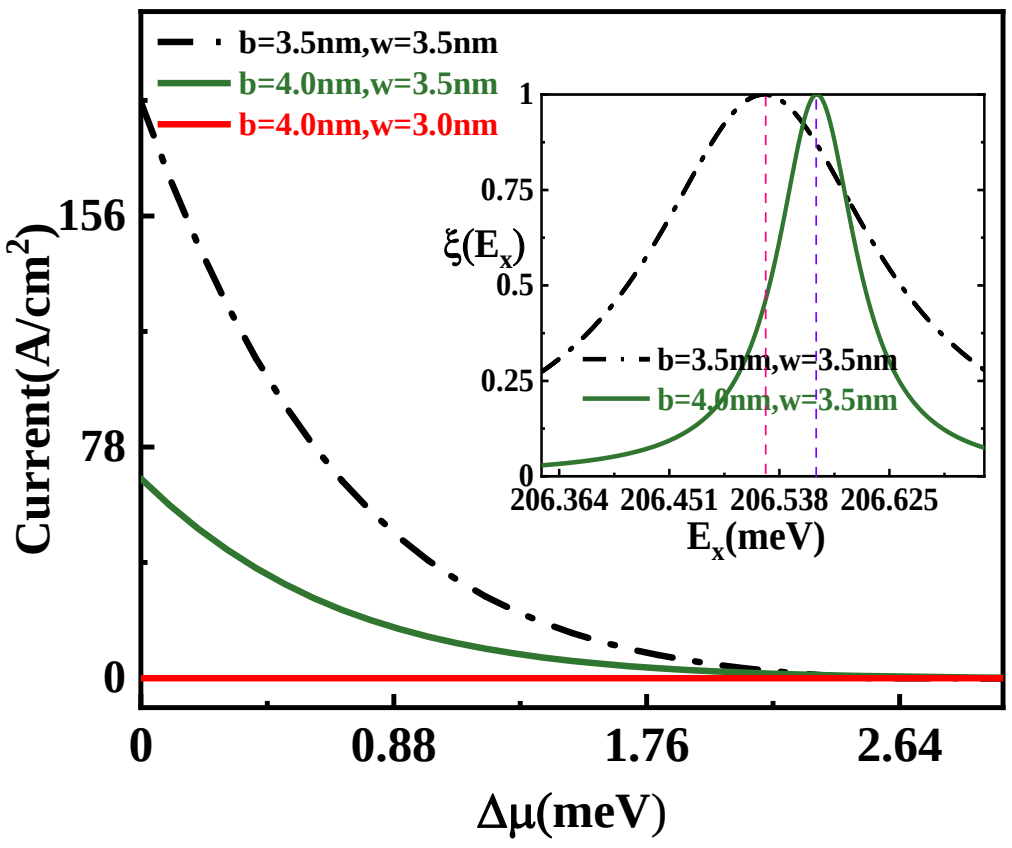

Fig.4. The net electron current density $J_{net}$ varying with $V_{bias}$ for different widths of barrier and well as labelled, where $\mu_C = 200\,\text{meV}$. The inserted figure shows the curves of $\xi(E_x)$ varying with $E_x$ for $b = 3.5\,\text{nm}, w = 3.5\,\text{nm}$ (black dash line) and $b = 4.0\,\text{nm}$, $w = 3.5\,\text{nm}$ (green solid line), where $E_x$ is in the range from $E_x = 206.3\,\text{meV}$ to $E_x = 206.7\,\text{meV}$.

We consider the heat leak from the hot reservoir to the cold one by the equation of phonon radiative transfer [37]

$$\dot{Q}_L = \sigma_P (T_H^4 - T_C^4)\,, \tag{8}$$

where $\sigma_P = 1.0 \times 10^{-6}\,\mathrm{W\,cm^{-2}\,K^{-4}}$ is the coefficient of the heat leak analogous to the Stefan–Boltzmann constant of phonon. The power output $P$ and efficiency $\eta$ of the thermoelectric device are, respectively, expressed as

$$P = J_{net} V_{bias} \tag{9}$$

and

$$\eta = P / (\dot{Q}_H + \dot{Q}_L)\,. \tag{10}$$

Fig. 5 gives the three-dimensional graphs of $P$ and $\eta$ varying with the chemical potential $\mu_C$ of the cold reservoir and the chemical potential difference $\Delta\mu = eV_{bias}$. For a given value of $\mu_C$, $P$ first increases as $\Delta\mu$ increases. After $P$ reaches a maximum value, the electronic flow $J_{net}$ is dramatically reduced [Fig. 4] as $\Delta\mu$ continues to increase, resulting in the decrease of $P$. Similarly, $P$ is not a monotonic function of $\mu_C$. It is observed from Fig. 5 (a) that when $\mu_C$ and $\Delta\mu$ are, respectively, equal to their respective optimal values $(\mu_C)_P$=$206.5\,\mathrm{meV}$ and $(\Delta\mu)_P = 0.60\,\mathrm{meV}$, the power output attains its local maximum. Fig.5 shows that the local maximum power output and efficiency are two different states. When $\mu_C$ and $\Delta\mu$ are, respectively, equal to their respective optimal values $(\mu_C)_\eta = 206.2\,\mathrm{meV}$ and $(\Delta\mu)_\eta = 0.67\,\mathrm{meV}$, the efficiency attains its local maximum. Fig. 5 also shows that $(\mu_C)_\eta < (\mu_C)_P$ and $(\Delta\mu)_\eta > (\Delta\mu)_P$.

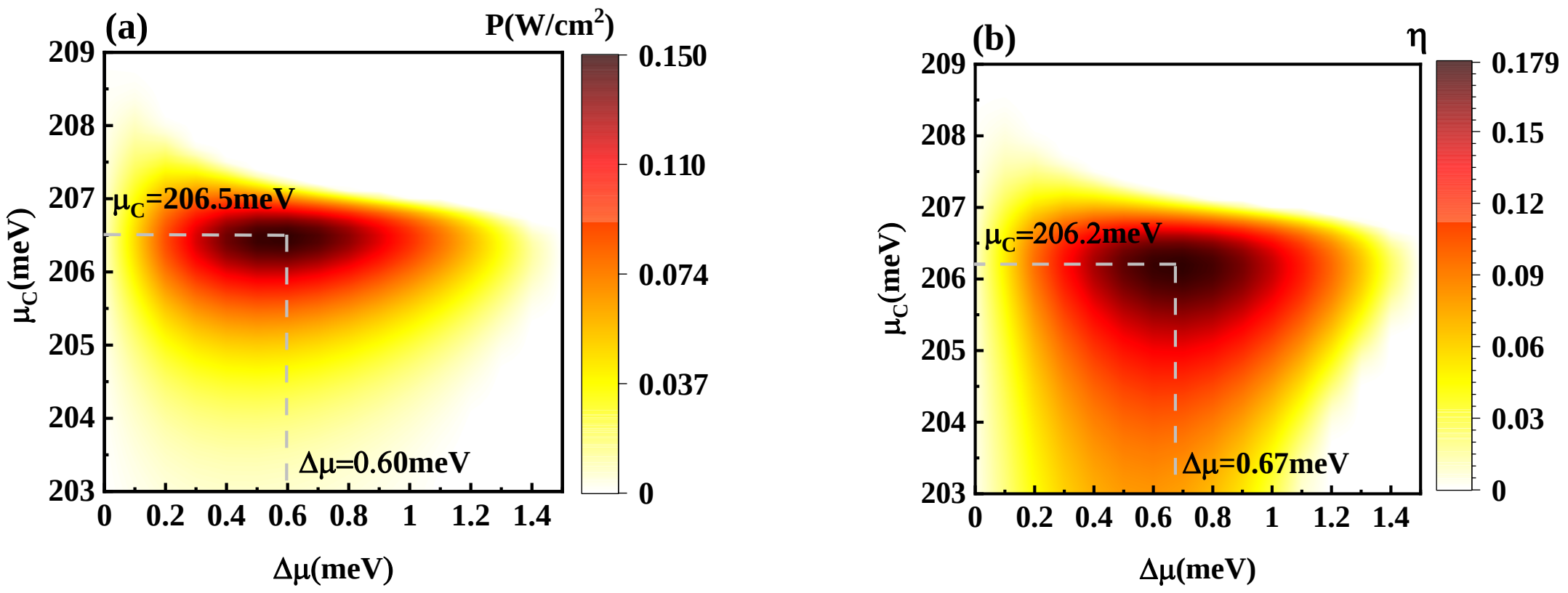

Fig.5. The three-dimensional graphs of (a) $P$ and (b) $\eta$ varying with the chemical potential $\mu_C$ of the cold reservoir and the chemical potential difference $\Delta\mu = eV_{bias}$, where $b = 3.0\,\mathrm{nm}$, and $w = 3.5\mathrm{nm}$.

For the given values of $\mu_C$ and $\Delta\mu$, one can also plot the three-dimensional graphs of $P$ and $\eta$ varying with the barrier width $b$ and well width $w$, as indicated by Figs. 6(a) and (b), respectively. It is seen from Fig. 6 that both $P$ and $\eta$ are not monotonical functions of $b$ and $w$, and the local maximum power output and efficiency are also two different states. When $b$ and $w$ are, respectively, equal to their respective optimal values $b_P = 2.752\,\mathrm{nm}$ and $w_P$=3.491nm, the power output attains its local maximum. When $b$ and $w$ are, respectively, equal to their respective optimal values $b_\eta$ and $w_\eta$, the efficiency attains its local maximum. It can be seen without difficulty that $b_\eta > b_P$ and $w_\eta > w_P$. The line connected by star in Fig. 6 (a) represents the stopping energy level, where the current driven by the temperature-gradient $\Delta T$ in the forward direction is compensated accurately by the bias driven current flowing in the opposite direction. The white region in the upper left corner of Fig. 6 (a) is not the operation regime of the heat engine, where the electron flows from the cold reservoir to the hot reservoir along the chemical potential difference $\Delta\mu$. In addition, in the working region of the heat engine, the power output increases initially and then decreases with the increase of barriers width $b$ and well width $w$, respectively. It should be noted that the power reduces and vanishes to zero with the decrease of well width and the increase of barrier width (lower right part of Fig.6 (a)). It can be explained by the fact that the decrease of well width and the increase of the barrier width also contribute to the resonance energy level $E_{\mathrm{res}}$ to shift to higher energy levels (along the direction of green arrow), where less electrons can be transmitted from the occupied state of the hot reservoir to the cold reservoir. In general, the heat engine cannot work at the maximum output power and maximum efficiency simultaneously. Thus, an optimally working region is required to trade off the efficiency and power. Fig. 6 (b) shows that the efficiency increases to its maximum and then decreases with respect to $b$ and $w$, and the local maximum efficiency is obtained at $b_\eta = 3.960\,\mathrm{nm}$, $w_\eta = 3.498\,\mathrm{nm}$.

When four parameters $\mu_C$, $\Delta\mu$, $b$, and $w$ are optimized simultaneously, one can obtain the characteristic curve of the efficiency versus power output, as shown in Fig.7, where $\eta_P$ is the efficiency at the maximum power output $P_{\max}$ and $P_\eta$ is the power output at the maximum efficiency $\eta_{\max}$. $P_{\max}$ and $\eta_{\max}$ give, respectively, the upper bounds of the power output and efficiency. For a nanostructured thermoelectric device, one always wants to obtain a high efficiency and a large power output as far as possible. Thus, according to Fig.7, the optimal regions of the power output and efficiency should be

$$P_\eta \le P \le P_{\max} \tag{11}$$

and

$$\eta_P \le \eta \le \eta_{\max}\,, \tag{12}$$

which correspond to the green curve with negative slope shown in Fig.7.

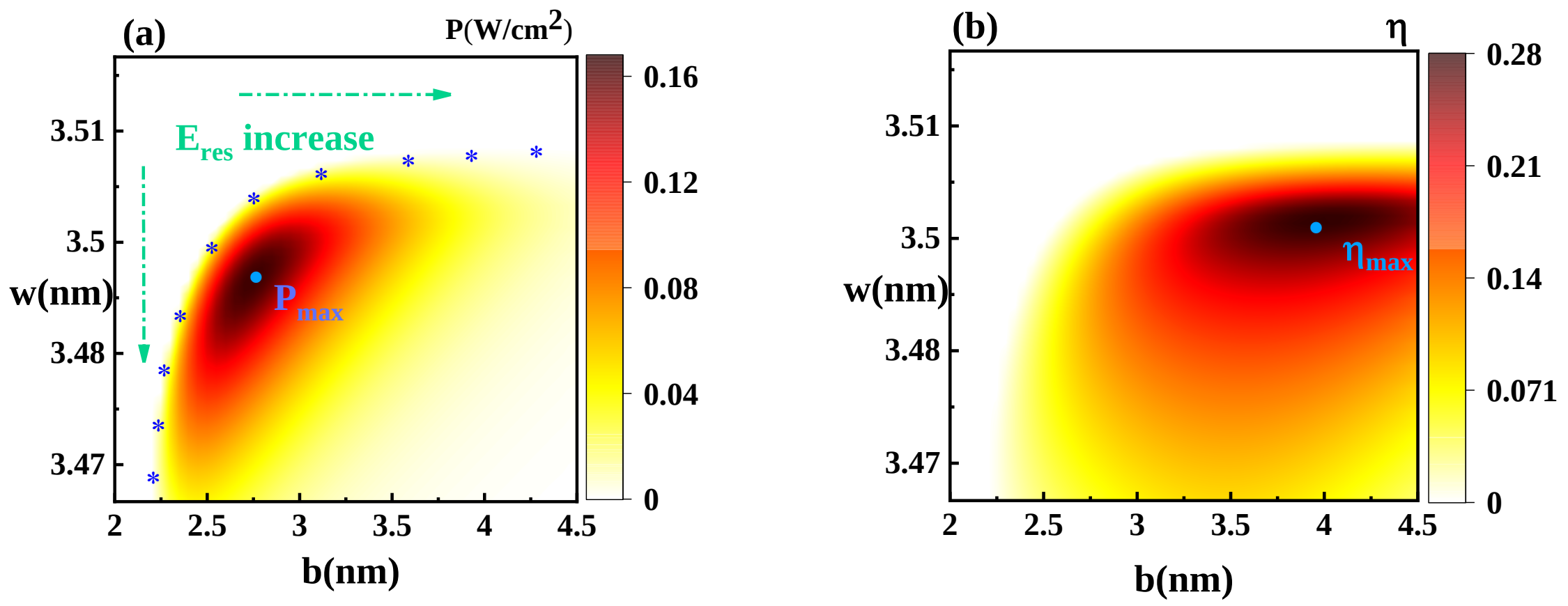

Fig.6. The (a) power output and (b) efficiency as functions of the barrier width $b$ and well width $w$, where $\mu_C = 206.9\,\mathrm{meV}$ and $\Delta\mu = 0.6\,\mathrm{meV}$.

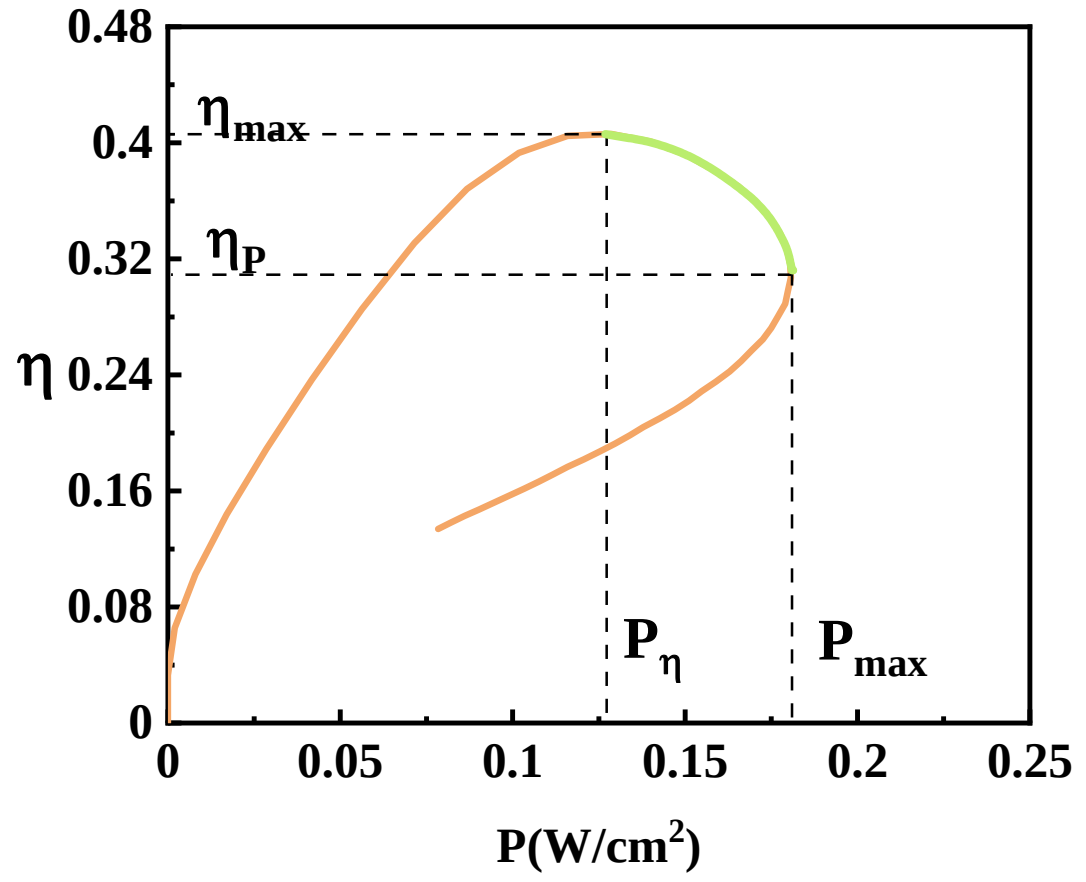

Fig.7. The optimum characteristic curve of the efficiency $\eta$ versus power output $P$.

According to Eqs. (11) and (12), one can directly determine the optimum regions of

other parameters as

$$\mu_{c,\eta} \le \mu_c \le \mu_{c,P}, \quad (13)$$

$$(\Delta\mu)_P \le \Delta\mu \le (\Delta\mu)_\eta \quad (14)$$

$$b_P \le b \le b_\eta, \quad (15)$$

and

$$w_P \le w \le w_\eta, \quad (16)$$

where $\mu_{c,P}$, $(\Delta\mu)_\eta$, $b_\eta$, and $w_\eta$ are the upper bounds of optimized parameters, and $\mu_{c,\eta}$, $(\Delta\mu)_P$, $b_P$, and $w_P$ are the lower bounds of optimized parameters. The maximum power output and efficiency and the boundary values of optimized parameters are listed in Table 1. Eqs. (11)-(16) and Table 1 may provide the optimal selection criteria for the main parameters of nanostructured thermoelectric devices.

Table 1 The optimal values of some parameters at the maximum power and efficiency

| $P_{\max}$ ($W/cm^2$) | $b_P$ (nm) | $w_P$ (nm) | $\mu_{C,P}$ (meV) | $(\Delta\mu)_P$ (meV) | $\eta_{\max}$ | $b_\eta$ (nm) | $w_\eta$ (nm) | $\mu_{C,\eta}$ (meV) | $(\Delta\mu)_\eta$ (meV) |
|---|---|---|---|---|---|---|---|---|---|
| 0.181 | 2.752 | 3.493 | 206.8 | 0.627 | 0.402 | 3.836 | 3.501 | 206.6 | 0.705 |

## 4. Conclusions

A one-dimensional double-barrier resonant tunneling heterojunction has been adopted to study the thermoelectric performance of a nanostructured device. By comparing the exact analytical solution with the solution obtained from the MB approximation, it is found that the method based on the MB approximation underestimates the magnitude of the heat flow. The transfer matrix method further shows that the electron and heat flows rely on the structure parameters of the heterojunction. The power output density and efficiency have been locally maximized by optimizing the chemical potential of the cold reservoir and the bias voltage for the given barrier and well widths or optimizing the barrier and well widths for the given chemical potential of the cold reservoir and the bias voltage. The optimum characteristic curve is obtained and the maximum power output density and efficiency are calculated. The optimally working regions of the thermoelectric device are determined, and the selection criteria of main parameters are supplied. These results obtained here may promote the experiment development of the nanostructured thermoelectric devices with resonance tunneling.

**ACKNOWLEDGMENTS**

This work has been supported by the National Natural Science Foundation of China (No. 11805159 and 12075197) and the Natural Science Foundation of Fujian Province of China (No. 2019J05003).

## Appendix A. The transfer matrix technique for the transmission probability

The resonant tunneling in the double-barrier heterostructure is a quantum effect in which the electron transmission probability $\xi(E_x)$ is sharply peaked at certain energies. $\xi(E_x)$ depends only on the energy of an electron in the transport direction and is restricted to $0 \le \xi(E_x) \le 1$. The chemical potential difference between the hot and cold reservoirs results in the asymmetric distribution of the potential, which increases the complexity for determining $\xi(E_x)$. By assuming that the voltage bias increases linearly from 0 at $x = 0$ to $V_{bias}$ at $x = 2b + w$, the transfer matrix method is applied to calculate $\xi(E_x)$. The solid line in the band diagram of Fig.1 (a) represents the potential profile $U(x)$ used in the calculation, which is given by

$$U(x) = \begin{cases} \phi + x\dfrac{\Delta\mu}{\lambda}, 0 < x < b \\ x\dfrac{\Delta\mu}{\lambda}, b < x < w + b \\ \phi + x\dfrac{\Delta\mu}{\lambda}, w + b < x < \lambda \end{cases}, \tag{A1}$$

where $\lambda$ is the total length of the double-barrier resonant tunneling heterostructure.

By solving the Schrodinger equation, the wave function outside the heterojunction has the following form [38]

$$\psi(x) = \begin{cases} Ae^{ik_0 x} + Be^{-ik_0 x}, & x < 0 \\ Ce^{-ik_0(x-\lambda)} + De^{-ik_0(x-\lambda)}, & x > \lambda \end{cases}, \tag{A2}$$

where $k_0 = \sqrt{2m^*_{GaAs}E_x} / \hbar$ is the wave vector of free electrons in the hot reservoir, $k = \sqrt{2m^*_{GaAs}(E_x + eV_{bias})} / \hbar$ is the counterpart of free electrons in the cold reservoir, $m^*_{GaAs}$ is the effective mass of electrons in reservoirs, and $A$ and $B$ are the coefficients of the forward traveling wave and reflection wave for electrons in the left side of the heterojunction, respectively. Similarly, $C$ and $D$ are the coefficients of the wave functions for electrons in the right hand side of the junction. At $x=0$ and $x=\lambda$, the wave function $\psi(x)$ and its derivative $\psi'(x)$ satisfy the continuity conditions. In this way, the amplitudes in Eq. (A2) can be rewritten in a matrix form

$$\begin{pmatrix} C \\ D \end{pmatrix} = p \begin{pmatrix} A \\ B \end{pmatrix}, \tag{A3}$$

where the matrix $p$ is given by

$$p = \frac{1}{2}\begin{pmatrix} 1 & -i/k \\ 1 & i/k \end{pmatrix} M(0,\lambda) \begin{pmatrix} 1 & 1 \\ ik_0 & -ik_0 \end{pmatrix}, \tag{A4}$$

with $M(0,\lambda)$ being the transfer matrix connecting the boundaries $x=0$ and $x=\lambda$. For a potential profile of arbitrary shape, it is convenient to compute $M(0,\lambda)$ by dividing the heterojunction into $N$ pieces with identical thickness $(2b+w)/N$. For $N\to\infty$, the potential in each piece can be regarded as a constant. As a result, $M(0,\lambda)$ can be decomposed as [39]

$$M(0,\lambda) = M_1 M_2 \cdots\cdots M_{N-1} M_N \,. \tag{A5}$$

If $0 \le (2b+w)\alpha/N \le b$ or $w+b \le (2b+w)\alpha/N \le \lambda$, piece $\alpha$ is located in the region of barriers with transfer matrix

$$M_\alpha = \begin{pmatrix} \cosh\kappa_\alpha b_\alpha & (\chi/\kappa_\alpha)\sinh\kappa_\alpha b_\alpha \\ (\kappa_\alpha/\chi)\sinh\kappa_\alpha b_\alpha & \cosh\kappa_\alpha b_\alpha \end{pmatrix}, \tag{A6}$$

where the wave vector $\kappa_\alpha = \sqrt{2m^*_{\mathrm{Al_xGa_{1-x}As}}\left\{E_x + U[(2b+w)\alpha/N]\right\}}/\hbar$ and $\chi$ is the ratio of the effective mass of electron in piece $\alpha$ to that of electron in piece $\alpha+1$. For $b \le (2b+w)\alpha/N \le b+w$, piece $\alpha$ is located in the region of the well with transfer matrix

$$M_\alpha = \begin{pmatrix} \cos k_\alpha w_\alpha & k_\alpha^{-1} \sin k_\alpha w_\alpha \\ -k_\alpha \sin k_\alpha w_\alpha & \cos k_\alpha w_\alpha \end{pmatrix}, \tag{A7}$$

where $\kappa_\alpha = \sqrt{2m^*_{\mathrm{GaAs}} \left\{ E_x + U[\left(2b+w\right)\alpha / N] \right\}} / \hbar$.

By combining Eqs. (A1)-(A8), the transmission probability $\xi(E_x)$ is given by

$$\xi(E_x) = \frac{|C|^2 k}{|A|^2 k_0} = \frac{4k_0 / k}{(S_{11} + k_0 S_{12} / k)^2 + (k_0 S_{12} - S_{21} / k)^2}. \tag{A8}$$

**References**

[1] Mateos J H, Real M A, Reichl C, et al. Thermoelectric cooling properties of a quantum Hall Corbino device. Phys Rev B, 2021, 103: 125404

[2] Da Y, Xuan Y M. Perfect solar absorber based on nanocone structured surface for high-efficiency solar thermoelectric generators. Sci China Tech Sci, 2015, 58(1): 19-28

[3] Mukherjee S, De B, Muralidharan B, Three-terminal vibron-coupled hybrid quantum dot thermoelectric refrigeration. J Appl Phys, 2020, 128: 234303

[4] Shu Y, Lin X, Qin H, et al. Quantum dots for display applications. Angew Chem Int Edit, 2020, 59(50): 22312-22323

[5] Li J, Tan L L, Chai S P. Heterojunction photocatalysts for artificial nitrogen fixation: fundamentals, latest advances and future perspectives. Nanoscale, 2021, 13: 7011-7033

[6] Nehra M, Dilbaghi N, Marrazza G, et al. 1D semiconductor nanowires for energy conversion, harvesting and storage applications. Nano Energy, 2020, 104991

[7] He J H, Guo H Z, Yue X K, et al. Magnetic field asymmetry of transport dynamics in a Coulomb-coupled quantum point contact. Phys Rev B, 2021, 103: 245309

[8] Josefsson M, Svilans A, Burke A M, et al. A quantum-dot heat engine operating close to the thermodynamic efficiency limits. Nat Nanotechnol, 2018, 13(10): 920-924

[9] Josefsson M, Svilans A, Linke H, et al. Optimal power and efficiency of single quantum dot heat engines: Theory and experiment. Phys Rev B, 2019, 99(23): 235432

[10] Kuo D M T. Thermoelectric and electron heat rectification properties of quantum dot superlattice nanowire arrays. AIP Adv, 2020, 10(4): 045222

[11] Nakpathomkun N, Xu H Q, Linke H. Thermoelectric efficiency at maximum power in low-dimensional systems. Phys. Rev. B, 2010, 82: 235428

[12] Sothmann B, Sánchez R, Jordan A N. Thermoelectric energy harvesting with quantum dots. Nanotechnology, 2015, 26(3): 032001

[13] Sánchez R, Sothmann B, Jordan A N, et al. Correlations of heat and charge currents in quantum-dot thermoelectric engines. New J Phys, 2013, 15(12): 125001

[14] Ortega-Piwonka I, Piro O, Figueiredo J, et al. Bursting and excitability in neuromorphic resonant tunneling diodes. Phys Rev Appl, 2021, 15: 034017

[15] Uskov A V, Khurgin J B, Protsenko I E, et al. Excitation of plasmonic nanoantennas by nonresonant and resonant electron tunneling. Nanoscale, 2016, 8(30): 14573-14579

[16] Su H, Shi Z C, He J Z. Optimal performance analysis of a three-terminal thermoelectric refrigerator with ideal tunneling quantum dots. Chinese Phys Lett, 2015, 32(10): 100501

[17] Lin Z, Yang Y Y, Li W, et al. Three-terminal refrigerator based on resonant-tunneling quantum wells. Phys Rev E, 2020, 101(2): 022117

[18] Li H, Li G. Analysis of ballistic transport in nanoscale devices by using an accelerated finite element contact block reduction approach. J Appl Phys, 2014, 116: 084501

[19] Yamamoto K, Masuda K, Uchida K, et al. Strain-induced enhancement of the Seebeck effect in magnetic tunneling junctions via interface resonant tunneling: Ab initio study. Phys Rev B, 2020, 101:094430

[20] Mishra S K, Kumar A, Kaushik C P, et al. An efficient tunable thermoelectric device based on n and p doped graphene superlattice heterostructures. Superlattices microstruct, 2020,

142: 106520

[21] Castro E D G, Rothmayr F, Krüger S, et al. Resonant tunneling of electrons in AlSb/GaInAsSb double barrier quantum wells. AIP Adv, 2020, 10: 055024

[22] Yang R Q. Electronic states and interband tunneling conditions in type-II quantum well heterostructures. J Appl Phys, 2020, 127 025705

[23] Marchewka M. Finite-difference method applied for eight-band kp model for $HgCd_xTe/HgTe$ quantum well. Int J Mod Phys B, 2017, 31(20): 1750137

[24] Li G. Transfer matrix approach to electron transport in monolayer $MoS_2/MoO_x$ heterostructures. Mater Res Express, 2018, 5(5): 055013

[25] Shi X L, Long H, Wu J Z, et al. Theoretical optimization of inhomogeneous broadening in InGaN/GaN MQWs to polariton splitting at low temperature. Superlattices Microstruct, 2019, 128: 151-156

[26] Yang Y Y, Xu S, Li W, et al. Optimal performance of three-terminal nanowire heat engine based on one-dimensional ballistic conductors. Phys Scr, 2020, 95(9): 095001

[27] Yang Y Y, Xu S, He J Z, et al. Three-terminal thermionic heat engine based on semiconductor heterostructures. Chinese Phys Lett, 2020, 37(12): 120502

[28] Su S, Liao T, Chen X, et al. Hot-carrier solar cells with quantum well and dot energy selective contacts. IEEE J Quant Electron, 2015, 51( 9): 4800208

[29] Godoy S. Semiclassical theory of quantum diffusion in 1D: A stochastic process for the Landauer equation. J Chem Phys, 1991, 94(9): 6214-6218

[30] Li Z Y, Lin L L. Photonic band structures solved by a plane-wave-based transfer-matrix method. Phys Rev E, 2003, 67(4): 046607

[31] Humphrey T E, Newbury R, Taylor R P, et al. Reversible quantum Brownian heat engines for electrons. Phys Rev Lett, 2002, 89: 116801

[32] Su S H, Guo J, Su G, et al. Performance optimum analysis and load matching of an energy selective electron heat engine. Energy, 2012, 44: 570

[33] O'Dwyer M F. Solid-state refrigeration and power generation using semiconductor nanostructures, 2007, University of Wollongong

[34] Luo X G, Liu N, He J Z, et al. Performance analysis of a tunneling thermoelectric heat engine with nano-scaled quantum well. Appl Phys A, 2014, 117(3): 1031-1039

[35] Yagi S, Hijikata Y, Okada Y, et al. Quantum well double barrier resonant tunneling structures for selective contacts of hot carrier solar cells. 37th IEEE Photovoltaic Specialists Conference, 2011, pp. 003309–003312

[36] Whitney R S. Most efficient quantum thermoelectric at finite power output. Phys Rev Lett, 2014, 112(13): 130601

[37] Klitsner T, VanCleve J E, Fischer H E, et al. Phonon radiative heat transfer and surface scattering. Phys Rev B, 1988, 38: 7576–7594

[38] Petersen R. Theoretical Investigation of the Resonant Tunneling Phenomena and its Applications in Resonant Tunneling Diodes. Mini-project 6th Semestre Nano-Physics Student, 2007, Aalborg University

[39] Li L, Study of metal-insulator-metal diodes for photodetection. (Doctoral dissertation, 2013, University of Dayton)